# The Future of Cybersecurity in Southeast Asia along the Maritime Silk Road

Roberto Dillon

James Cook University Singapore

roberto.dillon@jcu.edu.edu.au

## Abstract

Southeast Asian countries play a significant role in developing the 'One Belt and One Road' initiative, especially in constructing the '21st Century Maritime Silk Road'. They have a high demand for connectivity. The resulting extensive digitalization has brought unprecedented opportunities for trade and economic growth and new cybersecurity challenges that threaten the countries' integrity, security, and resilience in their digital transformation efforts. This paper proposes an analysis of the prospects of the cyber security industry and educational ecosystems in four Southeast Asian countries, namely Vietnam, Singapore, Malaysia, and Indonesia, which are along the Maritime Silk Road, by using two novel metrics: the "Cybersecurity Education Prospects Index" (CEPI) and the "Cybersecurity Industry Prospects Index" (CIPI). The CEPI evaluates the state of cybersecurity education by assessing the availability and quality of cybersecurity degrees together with their ability to attract new students. On the other hand, the CIPI measures the potential for the cybersecurity industry's growth and development by assessing the talent pool needed to build and sustain its growth. Ultimately, this study emphasizes the vital importance of a healthy cybersecurity ecosystem where education is responsible for supporting the industry to ensure the security and reliability of commercial operations in these countries against a complex and evolving cyber threat landscape.

## Introduction

As the global economy becomes increasingly interconnected, maritime commerce plays a pivotal role in facilitating international trade. With the advent of the modern Silk Road, represented by maritime routes connecting major economic hubs [1], across Vietnam, Malaysia, Singapore and Indonesia (see figure 1) the reliance on digital technologies for efficient and secure operations has grown significantly. However, this digital transformation has introduced new vulnerabilities, leading to the appearance of sophisticated cyber threats, including data breaches, ransomware attacks and more that can target commercial, industrial, government and private entities [2].

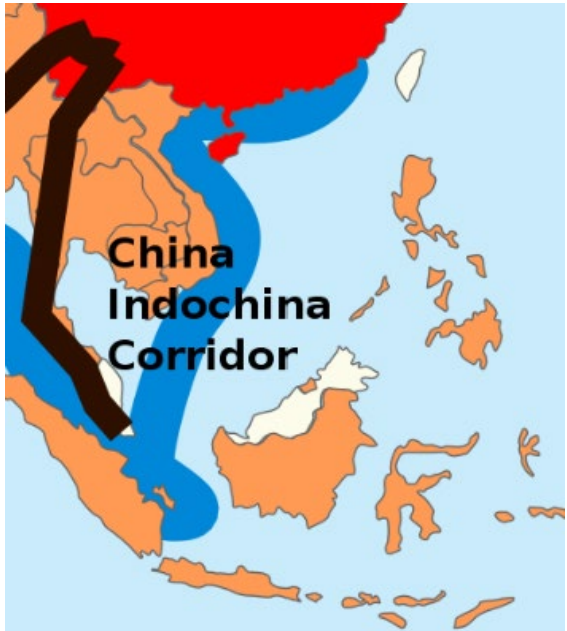

Figure 1: The Maritime Silk Road from China to Southeast Asia[1]

To improve our resilience across such a wide spectrum of threats, a close synergy between the cybersecurity industry, a field in constant need of qualified manpower ([3], [4]), and well recognized academic institutions able to supply such talent, is needed.

Which countries are then in a more favourable position to address these challenges?

## Evaluating the future of the cybersecurity ecosystem: the CEP and CIP Indexes

To answer the previous question, we have to assess a country ability to sustain the need for the industry to hire qualified professionals and, for the tertiary education system, to engage and educate new cohorts of students keen to enter a field that, while fascinating and rewarding, is also seen as very demanding, challenging and complex. As shown in Figure 2, Cybersecurity education is at the centre of a cybersecurity ecosystem, taking in input interested people and forming them so that they are ultimately able to join the workforce.

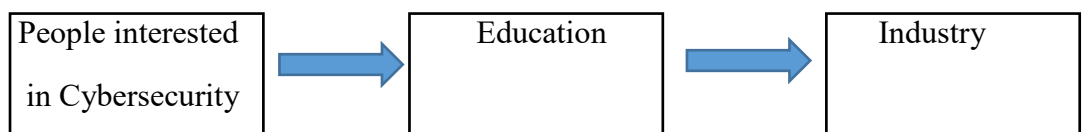

Figure 2: Education needs to constantly feed skilled professionals to the workforce to build a sustainable cybersecurity ecosystem.

---

[1] Belt and Road Initiative. (2023, July 24). In *Wikipedia*. https://en.wikipedia.org/wiki/Belt_and_Road_Initiative

To estimate the available pool of local talent in each country, i.e. young people within the school system and with a strong interest in cybersecurity, the metrics take into account two parameters: the tertiary education school enrolment percentage as reported by UNESCO [5] and the median value of the top 50 performers on the online training platform TryHackMe [6], respectively.

The former, named EP, for "Enrolment Percentage"[2] and shown in Table 1, can give us an idea of how accessible education is in a given country. This is a fundamental metric as a new cybersecurity workforce has to be highly educated to be proficient in this area.

Table 1: Tertiary school enrolment as a percentage of all eligible children (EP). World Average (2020): 40%

|    | Singapore | Malaysia | Indonesia | Vietnam |
|----|-----------|----------|-----------|---------|
| EP | 93        | 43       | 36        | 35      |

The latter instead, named THM, after the name of the platform, is important to gauge the interest on cybersecurity topics and education in the local communities. TryHackMe is one of the most popular training websites with, at the time of writing, more than 2.1 million users around the world. It offers a varied selection of gamified training activities and, by checking the score of the Top 50 performers in each country, we can have a measure of how many people are engaging with the platform (Table 2). The more competitive it is to reach the top of the leaderboard, the more people are engaged in learning activities and we can use this to extrapolate a general interest in cybersecurity. Note that the median value is used here instead of the average since we are more interested in measuring the performance of the overall community rather than in single outliers and top performers. A score higher than 20,000 is considered very high on the platform, giving users a rank of "God" and placing them in the Top 0.5% worldwide.

Table 2: Southeast Asian countries along the maritime silk road ranked by median score of their Top 50 THM performers (data taken on 15/7/23. Number of users are estimated via a simple linear regression model based on available data):

| # | Country   | THM Score (Top 50, Median) | Estimated Number of Users |
|---|-----------|----------------------------|---------------------------|
| 1 | Vietnam   | 32441.5                    | 17,000                    |
| 2 | Indonesia | 29308                      | 14,000                    |
| 3 | Malaysia  | 27346.5                    | 12,200                    |
| 4 | Singapore | 24461.5                    | 10,000                    |

For comparison's sake, Top 50 median score for the USA is 79,696.5 (world highest), for India it is 69,382, for Australia it is 38,114, for Egypt 33,683.5, for the UAE 23,410, for Sri Lanka 22,310, and for South Korea 11,957. All these values testify how global the interest in cybersecurity is and the raw potential that, in many countries, is still untapped.

---

[2] As defined by UNESCO: Gross enrolment ratio is the ratio of total enrolment, regardless of age, to the population of the age group that officially corresponds to the level of education shown.

To assess quality education opportunities, we can look at the dedicated cybersecurity degree programs, or related majors, being offered by QS ranked Universities in each country.

In particular, we have the following (Tables 3,4,5,6):

Table 3: Vietnam: 5 QS ranked Universities (2024). The following offer specialized degrees:

| University | QS Ranking | Degree |
|---|---|---|
| Duy Tan University | 514 | Bachelor of IT, Major in Network Security |
| Hanoi University of Science and Technology | 1201 | Bachelor in Cyber Security |
| | | Master in Computer Network and Information Security |

Table 4: Indonesia: 26 QS ranked Universities (2024). The following offer specialized degrees:

| University | QS Ranking | Degree |
|---|---|---|
| Bina Nusantara University | 1001 | Bachelor in Computing, Major in Cyber Security |
| Telkom University | 1001 | Master's Degree in Cyber Security and Digital Forensics |

Table 5: Malaysia: 28 QS ranked Universities (2024). The following offer specialized degrees:

| University | QS Ranking | Degree |
|---|---|---|
| Universiti Malaya | 65 | Master in Cyber Security |
| Universiti Putra Malaysia | 158 | Master in Information security |
| Universiti Kebangsaan Malaysia | 159 | Master in Cyber Security |
| Universiti Teknologi Malaysia | 188 | Bachelor of Computer Science, Major in Computer Networks & Security |
| Taylor's University | 284 | Bachelor of Computer Science (Hons), Major in Cyber Security |
| | | Master of Applied Computing, Major in Cybersecurity |
| Universiti Teknologi PETRONAS | 307 | Bachelor of IT (Hons), Major in Cyber Security |
| Universiti Utara Malaysia | 538 | Master of Science in Cybersecurity |
| Universiti Teknologi MARA - UiTM | 555 | Bachelor in Computer Science (Hons), Major in Computer Networks |
| | | Master Of Science In Cybersecurity And Digital Forensics |
| INTI International University | 556 | Bachelor in Computer Science (Hons), Major in Network and Security |
| Sunway University | 586 | Bachelor of Science (Hons) in IT, Major in Computer Networking and Security |
| Management and Science University | 621 | Bachelor (Hons) in Computer Forensic |
| Universiti Tenaga Nasional | 761 | Bachelor in Computer Science (Hons), Major in Cyber Security |
| Universiti Malaysia Pahang | 781 | Bachelor in Computer Science (Hons), Major in Cyber Security |
| Multimedia University | 1001 | Bachelor of Information Technology (Hons.), Major in Security Technology |

| University | QS Ranking | Degree |
|---|---|---|
| Universiti Kuala Lumpur | 1001 | Bachelor of Information Technology (Hons), Major in Computer System Security |
| Universiti Tun Hussein Onn Malaysia | 1001 | Bachelor in Computer Science (Hons), Major in Information Security |
| | | Master of Computer Science, Major in Information Security |
| Universiti Teknikal Malaysia Melaka | 1201 | Bachelor of Computer Science. Major in Computer Security |

Table 6: Singapore: 5 QS ranked Universities (2024). The following offer specialized degrees:

| University | QS Ranking | Degree |
|---|---|---|
| National University of Singapore | 8 | Bachelor of Technology in Computing, Major in Cybersecurity |
| Nanyang Technological University | 26 | Master of Science in Cyber Security |
| James Cook University | 415 | Bachelor of Cybersecurity |
| Singapore University of Technology and Design | 429 | Master of Science in Security by Design |
| | | Master of Science in Technology and Design (Major in Cybersecurity) |

To summarize the information in the previous tables, we can define two parameters corresponding to the overall number of dedicated degrees (DD) and degrees with specialised majors (SM), as shown in Table 7:

Table 7: Overall number of Dedicated Degrees (DD) and Specialized majors (SM) per country

|    | Indonesia | Malaysia | Singapore | Vietnam |
|---|---|---|---|---|
| DD | 1 | 6 | 3 | 2 |
| SM | 1 | 14 | 2 | 1 |

With these values, we can now define two new metrics that can help us understand the big picture and evaluate the future prospects of the cybersecurity ecosystem in each of these countries. The CEPI (Cybersecurity Education Prospects Index) and CIPI (Cybersecurity Industry Prospects Index) were first introduced in [7] and defined as:

$$\text{CEPI} = \frac{\text{THM} * \text{EP}}{K \,(1 + \text{DD} + \text{SM})} \qquad K=100 \qquad (1)$$

$$\text{CIPI} = \frac{\text{THM} \,(1 + K_1 * \text{DD} + K_2 * \text{SM})}{100 - \text{EP}} \qquad K_1 = 2 \,;\, K_2 = 1 \qquad (2)$$

To assess the health and growth opportunities of cybersecurity education, CEPI takes into consideration the THM median score and multiplies it for the enrolment percentage (EP) as a way to evaluate the amount of possible students who can potentially have an interest in accessing the courses. Then it divides it according to the number of degrees being offered scaled by a factor K, which we can set, for example, to 100 for simplicity. A higher CEPI value indicates that a country not only has a reliable talent pool to draw upon, but it also has potential to develop further by establishing new courses to educate the next generation of cybersecurity talent. A low value may imply that the education sector may not have room to grow, due to either a lack of prospective students, or a student population unable to access quality education, or an already fully developed and saturated educational offering.

CIPI, on the other hand, aims at evaluating the future prospects of the industry. To do so, it multiplies the THM median score for a factor that takes into consideration the number of degrees being offered, as well as assigning a different weight to dedicated degrees compared to majors in more general degrees ($K_1$=2 and $K_2$=1 respectively). Then it divides the result by taking into account the percentage of students who do not have access to tertiary education, i.e. countries that have a low enrolment percentage will find more difficult to sustain industry growth since fewer new talents will be formed by the local educational system. A country with a low CIPI value will find it challenging to sustain a quality cybersecurity industry as it won't be able to provide a constant stream of new and qualified professionals. Table 8 summarizes CEPI and CIPI values for the countries under analysis.

Table 8: CEPI and CIPI values for Southeast Asian countries along the maritime Silk Road.

|  | Indonesia | Malaysia | Singapore | Vietnam |
| --- | --- | --- | --- | --- |
| CEPI | 3,547 | 560 | 3,792 | 2,838 |
| CIPI | 1,832 | 12,954 | 31,450 | 2,995 |

Discussion

All countries show a strong community of highly interested and engaged cybersecurity enthusiasts, as testified by a high THM score. This means they are fertile ground for the development of a new generation of cybersecurity professionals, and the corresponding CEPI and CIPI values allow us to draw some additional insights.

Malaysia stands out for having the widest offering of specialized cybersecurity degrees in the region. This, together with its strong cybersecurity community, makes it one of the best placed countries to have a strong professional workforce in the near future (high CIPI). On the other hand, the educational landscape is already fully developed and further growth in this specific area is unlikely (low CEPI).

Singapore has both the highest CEPI and CIPI values. Being the technological hub of the region, this is not surprising and it is due to its very high EP value, meaning that no talent will be left behind as virtually all the local young population will have a chance to fully develop its talents thanks to a very strong educational system. Since the number of specialized degrees being offered is still relatively small, it is also still possible to have further growth in this area as well.

Indonesia's high CEPI shows very good potential for growth in the cybersecurity education field, due to a highly engaged community coupled with a lacking of dedicated cybersecurity degrees from its top universities. This, on the other hand, together with its EP value below the world average, makes it challenging for a robust local cybersecurity industry to emerge, as highlighted by the lowest CIPI in Table 8. The industry can only grow after the educational offer becomes stronger.

Last but not least, Vietnam. This country is particularly noteworthy for having the most enthusiastic and most numerous users on THM among the countries here discussed. This shows a huge future potential that is nonetheless hampered by its low EP value, which, despite having improved significantly in the last few years, is still below the world average. This negatively affects both its final CEPI and CIPI results. Until quality education does not become more easily accessible, local talents won't be able to grow, and the industry will find it very hard to develop as a consequence.

## Conclusions

From the previous analysis, we can conclude that the Southeast Asian countries situated along the Maritime Silk Road are well positioned to build a strong cybersecurity ecosystem in the near future. This will have a beneficial impact not only on their internal infrastructure but will also indirectly contribute to securing commercial routes close to their territorial spaces, like the maritime silk road, by providing, for example, more secure ports and communication channels.

The proposed CEPI and CIPI metrics make the success and future prospect of countries such as Singapore apparent, as well as highlighting the rapid development of the specialised educational offerings in Malaysia, which reached maturity in the cybersecurity field much faster than any other country in Southeast Asia. Nonetheless, we should not forget the challenges that the region still faces as a whole. While initiatives like THM clearly show there is a growing talent pool of passionate hobbyists, the access to quality education is still an issue in several countries, and this may slow down the overall expected development as quality degrees are needed to turn a hobbyist into a professional and foster industry growth in an organic and sustainable way.